\documentclass{aa}
\usepackage{graphicx}
\usepackage{txfonts}
\begin{document}
   \title{Absolute timing of the Crab Pulsar at optical wavelengths with STJs}

   \subtitle{}

   \author{T. Oosterbroek\inst{1}
          \and
          J.H.J. de Bruijne\inst{2}
          \and
          D. Martin\inst{1}
          \and
          P. Verhoeve\inst{1}
          \and
          M.A.C. Perryman\inst{2}
          \and
          C. Erd\inst{1}
          \and
          R. Schulz\inst{2}
          }

   \offprints{T. Oosterbroek}

   \institute{Science Payload and Advanced Concepts Office, ESA,
ESTEC, Postbus 299, 2200 AG Noordwijk, The Netherlands\\
              \email{toosterb@rssd.esa.int}
         \and
             Research and Scientific Support Department of ESA, ESTEC,
Postbus 299, 2200 AG Noordwijk, The Netherlands\\
             }

   \date{Received March 23, 2006; accepted May 24, 2006}


  \abstract
   {We have observed the Crab Pulsar in the optical with S-Cam, an
    instrument based on Superconducting Tunneling Junctions (STJs) 
    with $\mu$s time resolution.}
   {Our aim was to study the delay between the radio and optical
    pulse.}
   {The Crab Pulsar was observed three times over a time span of
    almost 7 years, on two different locations, using three different
    versions of the instrument, and using two different GPS units.}
   {We consistently find that the optical peak leads the radio peak by
    49$\pm$90, 254$\pm$170, and 291$\pm$100~$\mu$s. On assumption of a
    constant optical lead, the weighted-average value is
    $\sim$170~$\mu$s, or when rejecting (based on a perhaps 
    questionable radio ephemeris) the first measurement, 273$\pm$100~$\mu$s.}
   {}

   \keywords{ pulsars: individual (Crab Pulsar, PSR J0534+2200) }

   \maketitle
%

\section{Introduction}

Precise timing of pulsar light curves throughout the electromagnetic spectrum
is a powerful tool to constrain theories of the spatial distribution of
various emission regions. In recent years, it has become clear that the main
and secondary pulses of the Crab Pulsar (PSR J0534+2200) are not aligned
in time at different wavelengths. X-rays are leading the radio pulse by
reported values of 344$\pm$40~$\mu$s (Rots et al.\ \cite{rots:2004}; RXTE
data) and 280$\pm$40~$\mu$s (Kuiper et al.\ \cite{kuiper:2003}; INTEGRAL data)
and $\gamma$-rays are leading the radio pulse by 241$\pm$29~$\mu$s (Kuiper et
al.\ \cite{kuiper:2003}; EGRET data. The uncertainty in this value does not
include the EGRET absolute timing uncertainty of better than 100~$\mu$s). At
optical wavelengths, the observations present a less coherent picture. Sanwal
(1999) has reported a time shift of 140~$\mu$s (optical leading the
radio). The uncertainty in this value is 20~$\mu$s in the determination of the
optical peak and 75~$\mu$s in the radio ephemeris. Shearer et al.\
(\cite{shearer:2003}) have reported a lead of 100$\pm$20~$\mu$s for
simultaneous optical and radio observations of giant radio pulses. Golden et
al.\ (\cite{golden:2000}) have reported that the optical pulse {\it trails}
the radio pulse by $\sim$80$\pm$60~$\mu$s. Romani et al.\ (\cite{romani:2001})
conclude that the radio and optical peaks are coincident to better than
30~$\mu$s, but their error excludes the uncertainty of the radio ephemeris
(150~$\mu$s). The internal inconsistency of these results -- if we assume that
the optical-radio delay is constant -- has prompted us to look into this
matter in detail using recent observations in combination with earlier data.

\section{Observations}

Observations were obtained with S-Cam3 in November 2005 on the ESA
Optical Ground Station (OGS) telescope on Tenerife and with S-Cam2 in
October 2000 and S-Cam1 in February 1999 (see Perryman et al.\
\cite{perryman:1999}) on the WHT telescope on La Palma (Table
\ref{tab:log}).  All S-Cam instruments are based on Superconducting
Tunneling Junctions (STJs), which register individual photons with
high time resolution and moderate spectral resolution (see e.g.\
Martin et al.\ \cite{martin:2003}). Times are provided by a GPS unit
which yields an accuracy better than 1~$\mu$s with respect to UTC in
individual photon arrival times. The time resolution of the instrument
was 5~$\mu$s for S-Cam1 and S-Cam2 and 1~$\mu$s for S-Cam3.

\begin{table*}
\caption{Log of observations.}
\label{tab:log}
\centering
\begin{tabular}{c c c c c c c c c} 
\hline\hline
Instrument & Observation period (YYYY-MM-DD) & Exposure (s) & Observatory & GPS Instrument\\
\hline
S-Cam1 & 1999-02-02 -- 1999-02-04& 4800 & WHT La Palma (4.2 m) & HP 58503A\\
S-Cam2 & 2000-10-01 -- 2000-10-02& 3317 & WHT La Palma (4.2 m) & HP 58503A\\
S-Cam3 & 2005-11-12 & 2616 & OGS Tenerife (1 m) & Zyfer GPStarplus Model 565\\
\hline
\hline
\end{tabular}
\end{table*}

\section{Analysis and Results}

The time tags of all photons were converted from UTC to Barycentric
Dynamical Time (TDB) at the solar system barycenter (a process we
refer to as ``barycentering''). This process involves conversion from
UTC to TAI (by -- for our observations -- adding 32 leap seconds) and
subsequent conversion from TAI to TT by adding the constant offset of
32.184 s. TT is then converted TDB. (Although the IAU officially
replaced TDB by TCB in 1991, they reckoned that, where continuity with
previous work is desirable, TDB may be still be used.) Furthermore, a
correction from the observatory to the Earth centre and a correction
for the gravitational propagation delay are applied. We have used the
JPL DE200 ephemeris (and not the more recent DE405 ephemeris) to
perform the barycentering, since we compare our data against
observations of the Crab Pulsar in the radio, which are also
barycentered using this ephemeris. WGS84 coordinates of the telescopes
were obtained with a GPS unit. These geodetic coordinates were
converted into geocentric coordinates to obtain the proper observatory
to Earth centre correction.

\begin{figure}
\includegraphics[width=7cm,angle=-90]{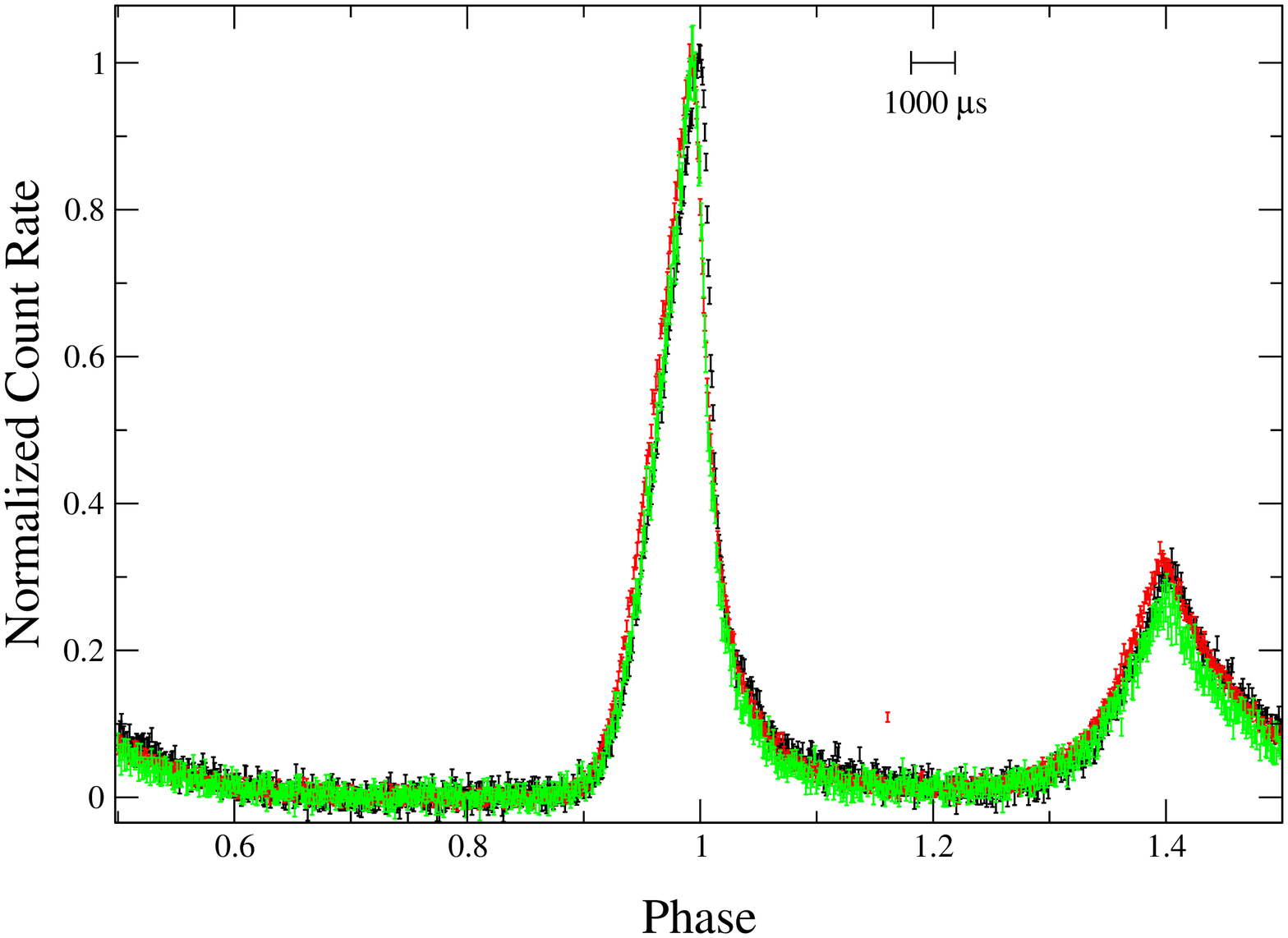}
\includegraphics[width=7cm,angle=-90]{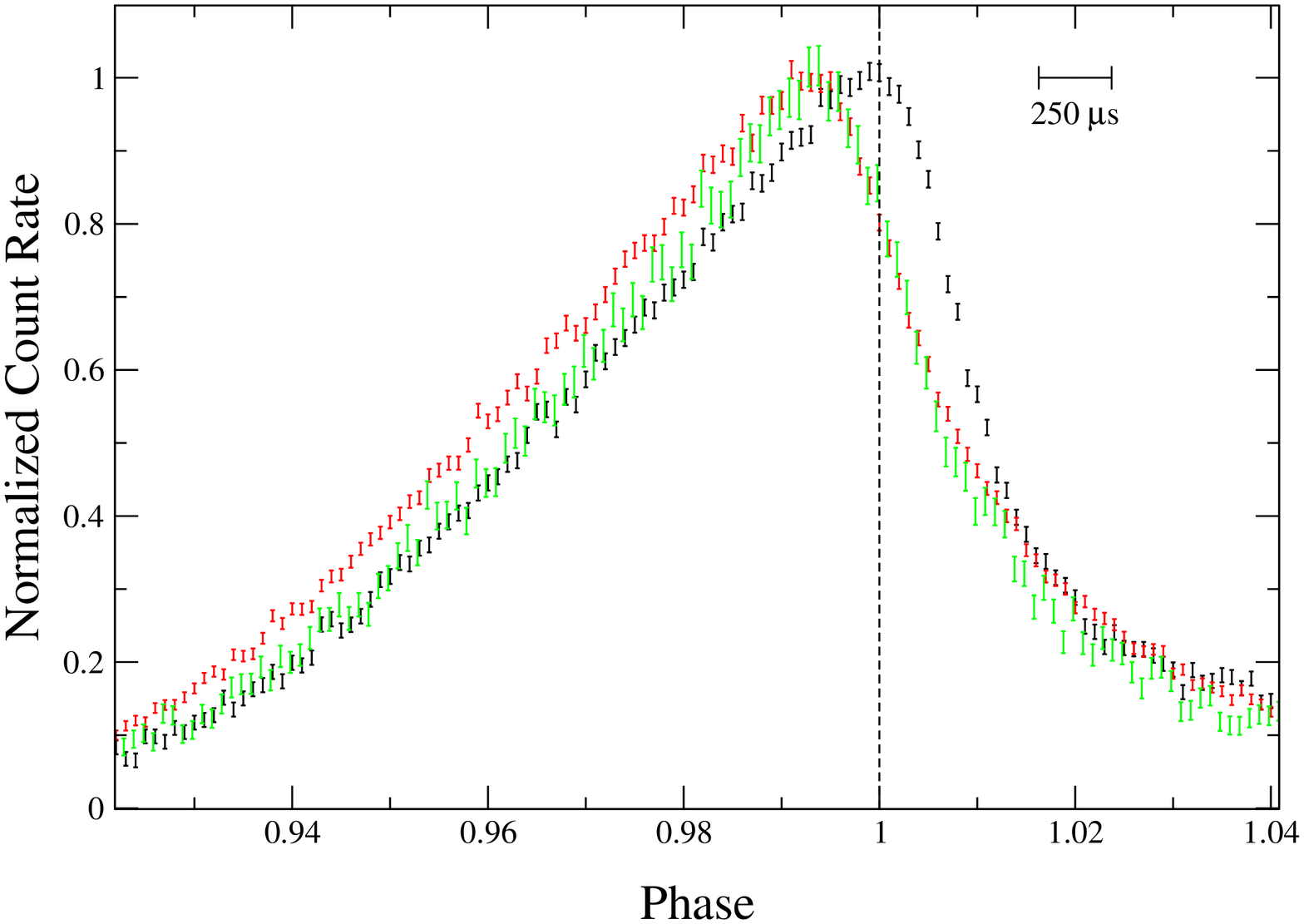}
\vspace{-1.5cm}
\caption[]{The three renormalized profiles: S-Cam1 in black, S-Cam2 in
  red, S-Cam3 in green (in the electronic version). In the top panel,
  the total profile is displayed. The bottom panel shows a zoom-in on
  the main peak. The statistical noise is largest for the S-Cam3
  observation, due to the smaller telescope and shorter exposure time.
  Phase 0 (or 1) is defined as the peak of the radio ephemeris; this
  phase is indicated by the vertical dashed line.}
\label{fig:profiles}
\end{figure}

We have folded our data using the $P$, $\dot{P}$, and arrival times (at
infinite frequency) from the Crab radio ephemeris (Lyne et
al. \cite{lyne:2005}). The folded profiles are displayed in Fig.\ 
\ref{fig:profiles}. An important parameter from this ephemeris is the
uncertainy in radio timing, which we denote with
$\sigma_{\mathrm{radio}}$. This uncertainty is composed of different elements
which can not all be treated as a statistical error that reduces when averaged
over multiple observations. The most important components are unmodeled delays
and imperfect polarization calibration ($\sim$40~$\mu$s), uncertainties in the
delay due to scattering ($\sim$20~$\mu$s), and variations in the daily
time-of-arrival measurements with rms residuals of the order 20--50~$\mu$s.

No default method exists for determining the position
of the peak of the profile. We follow the method of Rots et
al. (\cite{rots:2004}), which differs, however, from the choice of
other authors (e.g.\ Kuiper et al.  \cite{kuiper:2003}). The folded
pulse profile (subdivided into 1000 phase bins) was fitted in a phase
range of 0.01 centered around the peak using an iterative approach,
allowing the phase range in use to evolve with the fitted peak
position. The model used to determine the peak was a Lorentzian with
normalization, width, and center as free parameters. This choice is
somewhat arbitrary and different fitting functions (e.g.\ a Gaussian
or a parabola) return slightly different results. Variations, however,
are always much less than the uncertainty in the radio ephemeris, the
final number to which we compare against. We have also studied the
behaviour of the peak position obtained with different phase
ranges. These experiments show a consistent behaviour: due to the
asymmetric shape of the pulse profile, the phase of the peak shifts to
a higher value as the phase range is decreased but flattens off for
phase ranges smaller than 0.01. We estimate that the uncertainty
introduced by using a finite phase range of 0.01 (instead of an ideal,
infinitely small phase range) is 10~$\mu$s. Our results are summarised
in Table \ref{tab:delays}. In all cases, the optical pulse leads the
radio pulse. The time shift found from S-Cam1 data is clearly the
smallest, while the value from S-Cam2 data is affected by a
substantially larger $\sigma_{\mathrm{radio}}$.

We have also determined the position of the secondary peak using the
same algorithm as above but with twice the phase range (i.e. a width
of 0.02). This increased range is required since the secondary peak is
less sharp (resulting in almost no curvature in a 0.01 phase range)
and has a lower signal to noise ratio. The results (see Table
\ref{tab:delays}) are consistent with a constant phase difference
between the primary and secondary peak. This confirms that the S-Cam1
pulse profile is shifted (compared to the S-Cam2 and S-Cam3 data
points): not only the main peak arrives later, but also the secondary,
indicating a shift of the whole pulse profile, not a change in the
main peak only.

\begin{table*}
\caption{Radio to optical delays. The used parameters
  from the Crab monthly ephemeris are also displayed (converted into
  $P$ and $\dot{P}$). The number of digits displayed is higher than
  the uncertainty. The $\sigma_{\mathrm{radio}}$ quoted is the 
  uncertainty in the radio
  ephemeris as published in the Crab monthly ephemeris. The separation
  is defined as the difference (in phase) between the primary and
  secondary peak.}
\label{tab:delays}
\centering
\begin{tabular}{c c c c c c c c} 
\hline\hline
Instrument & Delay ($\mu$s) &  $\sigma_{\mathrm{radio}}$($\mu$s) & Separation &
                      $P$(s) & $\dot{P}$(10$^{-13}$s s$^{-1}$) & Epoch
                      (MJD)\\
\hline
S-Cam1 & 49$\pm$10 & 80 & 0.4056$\pm$0.0009  & 0.03349352792448 & 4.20554737 &
                      51224.000000352002\\
S-Cam2 & 254$\pm$8 & 160 & 0.4052$\pm$0.0005 & 0.03351561813489 & 4.20539152 &
                      51832.000000238553\\
S-Cam3 & 291$\pm$13 & 90 & 0.4065$\pm$0.0017& 0.03358309316492 & 4.20593301 &
                      53689.000000194479\\
\hline
\hline
\end{tabular}
\end{table*}

In order to compare the results of the three S-Cam campaigns, the
three different pulse profiles were scaled to match each other. The
level in the phase range 0.7--0.8 was calculated as well as
the peak level determined from the mean of the three highest bins. The
profile was renormalized such that the peak is at 1 and the base level
at 0. This scaling assumes that there are no significant counts from
the pulsar in the 0.7--0.8 phase range and that all other photons (sky
background, nebula background, and noise photons) are distributed
equally in phase.

The normalised profiles look similar, albeit the S-Cam1 profile looks
a bit different at the peak: the leading part of the peak is almost
identical to the S-Cam3 profile (up to phase 0.98), but then starts to
move away and peaks later than the S-Cam2 and S-Cam3 profiles. We have
investigated the possibility that the $P$ and $\dot{P}$ which have
been used to fold the data are sub-optimum by performing a period
search on our data. However, the solution hence found is fully
consistent with the radio ephemeris, with the latter having a higher
precision. Although we do not have evidence for a large
colour-dependence of the profile shape, small differences in the
detailed shape of the pulse profile may be explained by the different
spectral response and array size of the different generations of the
S-Cam instrument, possibly in combination with different
sky-background levels and atmospheric conditions under which the
observations were acquired.

We have also investigated, by exploiting the intrinsic spectral
resolution of STJs, whether there is a phase difference in the peak
between the ``red'' and the ``blue''. We have selected the $\sim$30\%
photons with highest energy ($\sim$400--500 nm) and the $\sim$30\%
photons with lowest energy ($\sim$600--760 nm). Because of the limited
spectral resolution of the S-Cam instrument, this approach yields a
cleaner colour separation than splitting the photons in two adjacent
bands, each containing 50\% of all events. We then determine the delay
between the photons in the red band with respect to the photons in the
blue band by doing a cross-correlation of the whole main peak
profile. We find that the red photons lag by $-$1$\pm$2~$\mu$s for
S-Cam1, 4$\pm$4~$\mu$s for S-Cam2, and 5$\pm$10~$\mu$s for S-Cam3.
This does not represent a significant time shift and can be compared
to the results of Sanwal (\cite{sanwal:1999}) who found a delay
between the U and R band of 10$\pm$4~$\mu$s and Golden et al.\
(\cite{golden:2000}) who found that the peaks in B and V are
coincident within 10~$\mu$s.


\section{Checks on correctness of times}

The following checks were made:

{\it GPS time:\/} It was verified that both GPS units provide the same
time to within 0.1~$\mu$s, the typical intrinsic GPS time jitter.

{\it GPS position:\/} The correct reading of the S-Cam3 GPS was
verified by the GPS present at the OGS telescope. This GPS presents
the position in Earth-Centered, Earth-Fixed (ECEF) coordinates (X,Y,Z)
= (5\,390\,284, $-$1\,597\,899, 3\,007\,075) m. The reading of the
S-Cam3 GPS (latitude 28$^{\circ}$ 18.$^{\mathrm{m}}$0640, longitude
16$^{\circ}$ 30.$^{\mathrm{m}}$7133 and altitude 2447.02 m) was
converted into ECEF coordinates which resulted in (X,Y,Z) =
(5\,390\,281, $-$1\,597\,891, 3\,007\,083) m. These coordinates differ
by 12 m, consistent with the separation between the two GPS
antennas.

{\it Period-folding software:\/} Two different approaches of folding
the data were followed and it was verified that they gave identical
results. The first approach is to fold the data with the period,
period derivative, and epoch as given by the Lyne et al.\ 
(\cite{lyne:2005}) radio-timing results. The second approach is to
determine, by phase folding, the ``current period'' (i.e.\ at the time
of the observation) and fold the data on that period without taking
the period derivative into account, which is not necessary for such
short datasets. The phase zero (or epoch) close to the time of the
observation should be independently calculated using the radio
ephemeris. While in theory both methods should give identical results,
this is by no means guaranteed in practice: since the time of
observation can be 15 days away from the ephemeris epoch, the
calculation of the phase using the period and period derivative
requires high precision. It was verified that this requirement on
precision is met by the {\sc FTOOLS} ``efold'' program.

{\it Barycentering software:\/} Using the detailed examples provided
in Lyne et al.\ (\cite{lyne:2005}), it was verified that our
barycentering code works correctly at the 5~$\mu$s level (dominated
by corrections in TT to TDB: relativistic corrections from the geoid
to the solar system barycenter).

{\it Crab coordinates:\/} For the barycentering of the Crab, the
coordinates used by Lyne et al.\ are: 05$^{\mathrm h}$ 34$^{\mathrm
m}$ 31.$^{\mathrm s}$97232, +22$^{\circ}$ 00$^{\prime}$
52.$^{\prime\prime}$0690 (J2000). For each monthly ephemeris, this
fixed position is used as an assumed position for the pulsar at the
relevant epoch. The known proper motion of $\sim$18 mas/yr is ignored,
and the effect of this proper motion on the true position is
incorporated into the solution of the period, period derivative and
arrival time (C.A.\ Jordan, private communication). Our barycentric
correction assumes the same reference position as the relevant monthly
ephemeris, so that consistent arrival times are obtained.  The
erroneous use of proper-motion corrected coordinates in the
barycentering would have a significant effect: inclusion of a proper
motion of 18 mas/yr would imply a time difference of $\sim$17~$\mu$sec
per year, or $\sim$100~$\mu$s for the 2005 observation.

{\it Instrumental effects:\/} The observation of delays/shifts at the
level of a few hundred $\mu$s and the realization that this time scale
is comparable to the average time between the events for the mean
count rate during the Crab observations, raises the possibility that
photons could have the time tag belonging to the previous photon. This
wrong association could either be caused by software or by
instrumental effects. Both possibilities were studied in detail and
rejected: we are confident that our photons have correct time tags.

{\it Instrumental effects (S-Cam2):\/} The S-Cam2 data suffered from a
hardware problem which resulted in corruption of time tags of 75\% of
all events, randomly distributed in time. Since this problem is fully
deterministic and understood, photons with corrupted time tags have
been filtered out. The remaining 25\% of photons, which have reliable
time tags, have been used for this analysis.

{\it Instrumental delays (S-Cam1 and S-Cam2):\/} In S-Cam1 and S-Cam2,
incoming photons generate electronic signals from which the photon
energy and arrival time can be extracted. The arrival time of a photon
is defined as the moment in time that the electronic signal rises
above a certain threshold. The delay of the registered arrival time
compared to the true arrival time is less than 10~$\mu$s, and
corrections for this delay have not been made for S-Cam1 and S-Cam2
data.

{\it Instrumental delays (S-Cam3):\/} In S-Cam3, each photon gives
rise to a bi-polar signal in the detector electronics chain. The
zero-crossing of this signal is used to time tag the photon. The delay
of this moment compared to the true time the photon entered the
detector has been calibrated to be 66.0$\pm$1.0~$\mu$s (fully
consistent with the value of 66.7~$\mu$s obtained from simulations of
the instrument). All S-Cam3 time tags have been corrected for this
(fixed) delay.

\section{Discussion and Conclusions}

\begin{figure}
\includegraphics[width=8cm,angle=0]{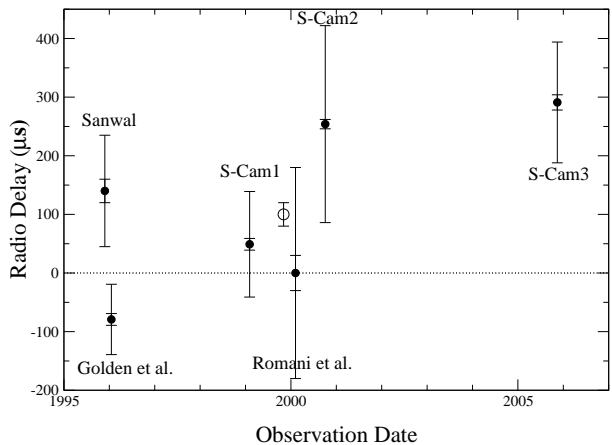}
\caption[]{Our three results combined with literature results in the
optical. For all results, the uncertainties were obtained by linear
addition (i.e.\ not quadratic) of the uncertainty in the peak
determination and the uncertainty in the radio
ephemeris. Uncertainties on the peak determination alone are also
plotted. The open symbol refers to Shearer et
al. (\cite{shearer:2003}), who observed giant radio pulses.}
\label{fig:results}
\end{figure}

We have observed the Crab Pulsar with three different generations of the S-Cam
instrument, on two locations, using two GPS units. We consistently find that
the optical pulse is leading the radio pulse.  However, the amount by which
the optical is leading the radio differs from observation to observation. When comparing the different results as plotted in Fig.\ \ref{fig:results}
which are obtained with the Jodrell Bank ephemeris, one should take into
account that $\sigma_{\mathrm{radio}}$ contains a systematical component (of
$\sim$40~$\mu$s) affecting all these measurements in the same way.If we
subtract the 40~$\mu$s, for the purpose of comparing the results, from
$\sigma_{\mathrm{radio}}$ and add the remainder, in quadrature, to our
measurement uncertainties, we obtain the following values for S-Cam1, S-Cam2,
and S-Cam3: 49$\pm$41, 254$\pm$120, and 291$\pm$50~$\mu$s.

From the last two observations (S-Cam2 and S-Cam3), we then determine an
average lead of 273$\pm$65~$\mu$s. The uncertainty in this determination
is dominated by uncertainties in the radio ephemeris. The result in
X-rays for the same two epochs is 370$\pm$40~$\mu$s (Rots, private
communication).


From our data, we cannot rigorously exclude the possibility that the
delay between the radio and optical peak evolves with time. The S-Cam1 result
(49$\pm$90~$\mu$s; the uncertainty has been determined by a linear addition of
the uncertainty in the peak determination and $\sigma_{\mathrm{radio}}$ ) in
particular is slightly deviant compared to the S-Cam2 and S-Cam3 data points
(254$\pm$170 and 291$\pm$100~$\mu$s).  However, the radio observations at the
times of the S-Cam1 observations have a high dispersion measure -- a higher
value has not been observed after February 1999. As a result, the
radio-ephemeris uncertainty (80~$\mu$s) might have been slightly
underestimated: variations in the dispersion measure could introduce a
somewhat higher uncertainty than the standard 20~$\mu$s (Lyne et al.\
\cite{lyne:1993}). We note that Rots et al.\ (\cite{rots:2004}) have
discarded their contemporaneous (X-ray) data point, because the timing
ephemeris is somewhat suspect (the GRO version has a high second derivative
indicative of a questionable fit). In the following we will therefore only
consider the S-Cam2 and S-Cam3 data points. We emphasize here that
corrections for arrival times at infinite frequency (which are substantial:
$\sim$0.6~s at 610 MHz) depend on the dispersion measure. The only
(literature) data point which is clearly inconsistent with all observations is
that of Golden et al.\ (\cite{golden:2000}). We have no explanation for this.

Our value of the optical phase difference between the main and
secondary pulse of 0.4054$\pm$0.0004 is not consistent with the X-ray
value from Rots et al.\ (\cite{rots:2004}) of 0.4001$\pm$0.0002
periods. This implies that the details of the pulse profile are
different in X-rays and at optical wavelengths.

Our time shift of 273$\pm$65~$\mu$s is somewhat smaller than, but
  consistent 
with, the time shift as obtained from X-ray measurements. A time shift of $\sim$270~$\mu$s indicates that possibly (in a simple geometrical model
ignoring relativistic effects) the optical radiation is formed 
$\sim$90~km higher in the magnetosphere than the radio
emission. Alternatively, the difference in phase of $\sim$0.008 could be
interpreted as an angle between the radio and optical beam of
$\sim$3$^{\circ}$.

Ideally, a simultaneous radio-optical observation at high frequency
should be performed. With an observation like this uncertainties
resulting from corrections for interstellar scattering are minimized,
and the accuracy will effectively be limited by systematic
effects ($\sim$40~$\mu$s for Jodrell Bank).

\begin{acknowledgements}
We thank Dr. D. Sanwal for providing details about his results. We thank the
referee, Dr A. Rots, for useful comments.
\end{acknowledgements}

\end{document}